\documentstyle[12pt]{article}

 \stop \fi

\batchmode
  \newfont{\footscrfont}{rsfs10}
  \newfont{\footbbbfont}{msbm10}
\errorstopmode

\newif\ifscrf\scrftrue
\ifx\footscrfont\nullfont
  \scrffalse
\fi

\newif\ifamsf\amsftrue
\ifx\footbbbfont\nullfont
  \amsffalse
\fi

\def\ppnumber{\vbox{\baselineskip14pt\hbox{CLNS-95/1334}
\hbox{hep-th/9505025}}}
\def\ppdate{May, 1995}
\def\pplogo{\vbox{\kern-\headheight\kern -15pt
\halign{##&##\hfil\cr&{
\ppnumber}\cr\rule{0pt}{2.5ex}&\ppdate\cr}
}}

\makeatletter
\date{}
\def\dedicatory#1{\def\@date{\normalsize\it#1}}
\def\subjclass#1{\def\@thefnmark{}\@footnotetext{1991
    {\it Mathematics Subject Classification.} #1}}
\def\keywords#1{\def\@thefnmark{}\@footnotetext{
    {\it Key words and phrases.} #1}}

\def\ps@firstpage{\ps@empty \def\@oddhead{\hss\pplogo}%
  \let\@evenhead\@oddhead 
}
\def\maketitle{\par
 \begingroup
 \def\thefootnote{\fnsymbol{footnote}}
 \def\@makefnmark{\hbox
 to 0pt{$^{\@thefnmark}$\hss}}
 \if@twocolumn
 \twocolumn[\@maketitle]
 \else \newpage
 \global\@topnum\z@ \@maketitle \fi\thispagestyle{firstpage}\@thanks
 \endgroup
 \setcounter{footnote}{0}
 \let\maketitle\relax
 \let\@maketitle\relax
 \gdef\@thanks{}\gdef\@author{}\gdef\@title{}\let\thanks\relax}

\def\abstract{\if@twocolumn
\section*{Abstract}
\else \small
\begin{center}
{\bf ABSTRACT}
\end{center}
\quotation
\fi}

\def\thebibliography#1{\section*{References\@mkboth
 {REFERENCES}{REFERENCES}}\small\list
 {[\arabic{enumi}]}{\settowidth\labelwidth{[#1]}\leftmargin\labelwidth
 \advance\leftmargin\labelsep
 \usecounter{enumi}}
 \def\newblock{\hskip .11em plus .33em minus .07em}
 \sloppy\clubpenalty4000\widowpenalty4000
 \sfcode`\.=1000\relax}

\newif\iffn\fnfalse

\@ifundefined{reset@font}{\let\reset@font\empty}{} 
\long\def\@footnotetext#1{\insert\footins{\reset@font\footnotesize
    \interlinepenalty\interfootnotelinepenalty
    \splittopskip\footnotesep
    \splitmaxdepth \dp\strutbox \floatingpenalty \@MM
    \hsize\columnwidth \@parboxrestore
   \edef\@currentlabel{\csname p@footnote\endcsname\@thefnmark}\@makefntext
    {\rule{\z@}{\footnotesep}\ignorespaces
      \fntrue#1\fnfalse\strut}}}
\makeatother

\ifamsf
  \newfont{\bigbbbfont}{msbm10 scaled\magstep2}
  \newfont{\bbbfont}{msbm10 scaled\magstep1}  
  \newfont{\smallbbbfont}{msbm8}
  \newfont{\tinybbbfont}{msbm6}
  \newfont{\smallfootbbbfont}{msbm7}
  \newfont{\tinyfootbbbfont}{msbm5}
\fi

\ifscrf
  \newfont{\scrfont}{rsfs10 scaled\magstep1}  
  \newfont{\smallscrfont}{rsfs7}
  \newfont{\tinyscrfont}{rsfs7}
  \newfont{\smallfootscrfont}{rsfs7}
  \newfont{\tinyfootscrfont}{rsfs7}
\fi

\ifamsf
  \newcommand{\Bbb}[1]{\iffn
      \mathchoice{\mbox{\footbbbfont #1}}{\mbox{\footbbbfont #1}}
      {\mbox{\smallfootbbbfont #1}}{\mbox{\tinyfootbbbfont #1}}\else
      \mathchoice{\mbox{\bbbfont #1}}{\mbox{\bbbfont #1}}
      {\mbox{\smallbbbfont #1}}{\mbox{\tinybbbfont #1}}\fi}
\else
  \def\bigbbbfont{\bf}
  \def\Bbb{\bf}
\fi

\ifscrf
  \newcommand{\Scr}[1]{\iffn
    \mathchoice{\mbox{\footscrfont #1}}{\mbox{\footscrfont #1}}
    {\mbox{\smallfootscrfont #1}}{\mbox{\tinyfootscrfont #1}}\else
    \mathchoice{\mbox{\scrfont #1}}{\mbox{\scrfont #1}}
    {\mbox{\smallscrfont #1}}{\mbox{\tinyscrfont #1}}\fi}
\else
  \def\Scr{\cal}
\fi

\def\operatorname#1{\mathop{\rm #1}\nolimits}

\def\R{{\Bbb R}}
\def\Z{{\Bbb Z}}

\def\Vol{\operatorname{Vol}}

\def\opeq#1{\advance\lineskip#1 \advance\baselineskip#1
        \advance\lineskiplimit#1}

\def\sm{$\sigma$-model}

\def\CY{Calabi--Yau}

\def\cM{{\Scr M}}

\def\cD{{\Scr D}}

\def\cMc{{\hfuzz=100cm\hbox to 0pt{$\;\overline{\phantom{X}}$}\cM}}
\def\barcD{{\hfuzz=100cm\hbox to 0pt{$\;\overline{\phantom{X}}$}\cD}}

\makeatletter

\relax
\@writefile{toc}{\string\contentsline\space {section}{\string\numberline%
\space {1}Introduction}{1}}
\@writefile{toc}{\string\contentsline\space {section}{\string\numberline%
\space {2}U-duality for K3 Surfaces}{2}}
\@writefile{toc}{\string\contentsline\space {section}{\string\numberline%
\space {3}The Case {\string\prm\space K3}$\times T^2$}{5}}
\@writefile{toc}{\string\contentsline\space {section}{\string\numberline%
\space {4}Exotic Limits}{7}}
\@writefile{toc}{\string\contentsline\space {section}{\string\numberline%
\space {5}Integral Structures}{9}}
\makeatother

\begin{document}
\setcounter{page}0
\title{\LARGE U-Duality and Integral Structures\\[10mm]}
\author{
Paul S. Aspinwall\\[0.7cm]
\normalsize F.R.~Newman Lab.~of Nuclear Studies,\\
\normalsize Cornell University,\\
\normalsize Ithaca, NY 14853\\[10mm]
David R. Morrison\thanks{On leave from:
Department of Mathematics, Duke University, Durham, NC 27708-0320}\\[0.7cm]
\normalsize Department of Mathematics, \\
\normalsize Cornell University, \\
\normalsize Ithaca, NY 14853\\[10mm]
}

{\hfuzz=10cm\maketitle}

\def\Large{\large}
\def\LARGE{\large\bf}

\begin{abstract}

We analyze the U-duality group for the case of a type II superstring
compactified to four dimensions on a K3 surface times a torus. The
various limits of this theory are considered which have
interpretations as type IIA and IIB superstrings, the heterotic string,
and eleven-dimensional supergravity, allowing all these theories to be
directly related to each other. The integral structure which
appears in the Ramond-Ramond sector of the type II superstring is
related to the quantum cohomology of general Calabi--Yau threefolds
which allows
the moduli space of type II superstring compactifications on \CY\
manifolds to be analyzed.

\end{abstract}

\vfil\break

\section{Introduction}          \label{s:intro}

\def\com#1{#1_c}

Consider a theory of supergravity in $d$ dimensions with $N$
supersymmetries. Given $d$, for a suitably large $N$ one expects the
local geometry of the  space parametrized by
the massless scalars to be tightly constrained. This was shown, for
example, for the case of maximal $N$ in \cite{CJ:SO(8),Jul:dis} and
for $N=4,d=4$ in \cite{deRoo:}. In each case the parameter space is of
the form $G/\com{G}$ where $\com{G}$ is the maximal compact subgroup
of the Lie group $G$.

If such a theory is derived as the low energy limit of a superstring
theory one generally expects that different values of the
parameters derive from inequivalent string vacua. However, some parameter
values
may correspond to equivalent theories. One therefore expects that the
moduli space of such string theories takes the form $U\backslash
G/\com{G}$, where $U$ is some discrete subgroup of $G$.
The group $U$ is a purely
stringy phenomenon and cannot be determined from considerations of
supergravity alone.

If one envisions the effective theory of supergravity to have arisen
as the compactification of some string theory on some compact space
$X$, then some of the elements of $U$ may be identified. At the
simplest level, the classical description of the moduli space of $X$
will lead to some of these identifications. The
``T-duality'' elements arise from a conformal field theory description of $X$
--- two geometrically distinct $X$'s may give rise to the same
conformal field theory. The ``S-duality'' elements arise from an
$Sl(2,\Z)$ acting on the axion-dilaton system (in ten dimensions)
and include a $\Z_2$
element that exchanges strong and weak string-coupling. It would be
unreasonable to expect conformal field theory to ``see'' such a
symmetry directly. In addition,
the S-duality and T-duality generators need not
commute (as was shown in the example studied in \cite{Sen:3d}, for
example) and so may generate a larger group. In the cases of interest
to us this larger group will be $U$, the group of ``U-dualities''.

The group $U$ was first analyzed in \cite{HT:unity} by making use of some
conjectures regarding the spectrum of solitons in the theory.
In this letter we will be concerned with U-duality in four-dimensional
theories. In particular we will focus on the case of the type II (A or B)
superstring compactified on K3 times a torus.
Rather than use the approach of \cite{HT:unity}, we will generally try
to avoid using any results requiring a knowledge of solitons (although
there has been some recent progress in this subject
\cite{CDFvP:,Sen:sol,HS:sol}).
Instead, we
will follow more closely the logic of using the simpler structure of various
weak coupling limits as was done in \cite{W:dyn}.
We examine how the discrete group $U$
may be derived purely from a knowledge of conformal field theory (and
hence T-duality) and the hypothesized S-duality on the type IIB
superstring in ten dimensions.

Unfortunately, for realistic models the value of $N$ is smaller
than what is  required
for the above scenario to work. In these cases the T-duality of conformal
field theory moduli space does not provide as much information
 and a phase structure
becomes important \cite{W:phase,AGM:II}. We must therefore
expect the U-duality picture to become much more subtle. However,
the breakdown
of the T-duality method has not prevented
the moduli spaces of general $N=2$ superconformal field theories from being
determined to a large extent.
In many cases mirror symmetry can effectively solve the
problem as was done in \cite{CDGP:} (for a review see
\cite{me:N2lect}). We will see that the integral structures coming
from the U-duality picture of the simple cases analyzed generalizes
naturally to the general \CY\ manifold and this likewise allows for
a determination
of the moduli space of type II superstring compactifications.


\section{U-duality for K3 Surfaces}

In order to explain the tools we will use for the desired case of a K3
surface times a torus we will explicitly review the simpler case of a type
IIB superstring compactified on a K3 surface \cite{W:dyn}.

Given a compactified string theory, one might hope to find
a more conventional physical description of the theory in a limit
where  some parameter tends to zero or
infinity, i.e., by going out to the boundary of the moduli space. This
idea was analyzed and applied in \cite{W:dyn} and we review the method
quickly here.

Consider some semi-simple Lie group $G$. Take the Dynkin diagram for
the algebra associated to the group and remove one vertex. The
resultant diagram may be associated to a subgroup $H$ of $G$. This
procedure determines a maximal embedding
\begin{equation}
  G \supset H\times J,  \label{eq:embed}
\end{equation}
where $J$ is a continuous abelian group of dimension 1. Clearly $J$
must be $U(1)$ or the group $\R_+$ of positive real numbers under
multiplication. We will be interested in the case
$J\cong\R_+$. For a given real form of a Lie group one may label the
dots in the Dynkin diagram according to whether the resultant $J$ will
be compact or not. For the case studied in \cite{W:dyn}, $G\cong
E_{7(7)}$ (which is a maximally noncompact form) and any dot leads to
$J\cong\R_+$.

Dividing both sides of (\ref{eq:embed}) in the case $J\cong\R_+$ by the
maximal compact subgroup leads in general to a decomposition of the following
sort
\begin{equation} \label{eq:gendecomp}
\frac{G}{\com{G}} \cong \frac{H}{\com{H}} \times \R_+ \times \R^M ,
\end{equation}
where
$\R^M$ is some linear vector space which comes equipped\footnote{In fact,
the subgroup of $G$ which preserves the decomposition (\ref{eq:gendecomp})
contains both $H$ and a translation subgroup isomorphic to  $\R^M$;
the representation of $H$ on $\R^M$ intertwines the two.}
 with a representation
of $H$.  The parameter describing $\R_+$ can be taken to
infinity to define the boundary required. Thus each dot associated to
a noncompact $J$ in the Dynkin diagram defines some limit in the
moduli space.

Compactifying the type IIB string on K3 to $d=6$ gives the
supergravity studied in \cite{Rom:6} leading to the result that
$G\cong O(5,21)$. This has the following Dynkin diagram
\begin{equation}
\setlength{\unitlength}{0.007in}%
\begin{picture}(450,70)(95,655)
\put(535,670){\makebox(0,0)[lb]{$e_1$}}
\put(495,670){\makebox(0,0)[lb]{$e_2$}}
\put(455,670){\makebox(0,0)[lb]{$e_3$}}
\put(415,670){\makebox(0,0)[lb]{$e_4$}}
\put(375,670){\makebox(0,0)[lb]{$e_5$}}
\thinlines
\put(140,690){\circle{10}}
\put(180,690){\circle{10}}
\put(220,690){\circle{10}}
\put(260,690){\circle{10}}
\put(300,690){\circle{10}}
\put(340,690){\circle{10}}
\put(540,690){\circle*{10}}
\put(380,690){\circle*{10}}
\put(420,690){\circle*{10}}
\put(100,720){\circle{10}}
\put(500,690){\circle*{10}}
\put(100,660){\circle{10}}
\put(460,690){\circle*{10}}
\put(145,690){\line( 1, 0){ 30}}
\put(185,690){\line( 1, 0){ 30}}
\put(225,690){\line( 1, 0){ 30}}
\put(265,690){\line( 1, 0){ 30}}
\put(305,690){\line( 1, 0){ 30}}
\put(345,690){\line( 1, 0){195}}
\put(104,716){\line( 4,-3){ 32}}
\put(104,663){\line( 4, 3){ 32}}
\end{picture}   \label{eq:D13}
\end{equation}
where the black dots represent roots associated to noncompact
generators. Thus we may obtain 5 limits by removing each of
$e_1,\ldots,e_5$ in turn. First, in order to make contact with the geometry
we understand, we should go to the large radius limit of the compact K3
surface. The corresponds to removing $e_2$ giving the following
decomposition
\begin{equation}
\frac{O(5,21)}{O(5)\times O(21)}\cong \frac{O(3,19)}{O(3)\times O(19)}
\times \frac{Sl(2)}{U(1)} \times \R_+ \times \R^{45}.   \label{eq:K3Bv}
\end{equation}
Up to discrete group identifications, the first factor on the
right-hand side is the moduli space of
Ricci-flat metrics on a K3 surface of fixed volume \cite{Beau:K3}
(including orbifold metrics \cite{Mor:Katata,Kobayashi-Todorov}).
The second factor is the moduli space of
type IIB strings in ten dimensions and the $\R_+$ is the volume of the
K3 surface. The 45 other fields consist of 22 $B$-field deformations and
23 R-R moduli. (There is one additional R-R field within the
$Sl(2)/U(1)$ factor.)
We may use this knowledge as follows to build up part of the
U-duality group on the left-hand side.

The group $O(5,21)$ acts naturally on the space
$\R^{5,21}$ into which we can embed a unique even self-dual lattice
$\Lambda^{5,21}$. The subgroup $O(\Lambda^{5,21})\subset O(5,21)$
preserving this lattice is a maximal discrete group \cite{Allan:}.
That is to say, if
we divide $O(5,21)$ by a group containing $O(\Lambda^{5,21})$ as a
proper subgroup the resulting quotient will not be Hausdorff. We will
embed discrete group actions on the right-hand side of (\ref{eq:K3Bv})
into $O(\Lambda^{5,21})$ and see how much of $O(\Lambda^{5,21})$ we
can generate.

The inner product on $\R^{5,21}$ may be applied to the generators of
$\Lambda^{5,21}$ to form the matrix $(-{\bf E}_8)^{\oplus2}\oplus{\bf H}
^{\oplus5}$, where ${\bf E}_8$ is the Cartan matrix of $E_8$ and
\begin{equation}
{\bf H} = \left(\begin{array}{cc}0&1\\1&0\end{array}\right). \label{eq:H}
\end{equation}

The discrete group action by which $O(3,19)/(O(3)\times O(19))$ should
be divided to obtain the moduli space of Ricci-flat metrics
is\footnote{In order to avoid cluttering notation we are being
a little careless with some $\Z_2$ factors. If these are properly
taken into account, our results are not affected.}
$O(\Lambda^{3,19})$. In this action, $\Lambda^{3,19}$
comes from the second cohomology group $H^2({ K3},\Z)$ and
has intersection form $(-{\bf E}_8)^{\oplus2}\oplus{\bf H}
^{\oplus3}$. It is clear how we embed $O(\Lambda^{3,19})$ into
$O(\Lambda^{5,21})$.

The conjecture for U-duality for the type IIB superstring in ten
dimensions tells us that the group $Sl(2)/U(1)$ should be divided by
$Sl(2,\Z)$ in the second factor of the right-hand side of
(\ref{eq:K3Bv}). Since  this part of the discrete group acts on
 a separate factor of the
decomposition, we must embed this $Sl(2,\Z)$  into
$O(\Lambda^{5,21})$ in such a way that it commutes with
$O(\Lambda^{3,19})$. This can be done as follows. Take the lattice
orthogonal to $\Lambda^{3,19}\subset \Lambda^{5,21}$ with
intersection form ${\bf H}\oplus{\bf H}$. Let $\alpha_1,\alpha_2$ be
two generators of this lattice such that
$\langle\alpha_1,\alpha_1\rangle=\langle\alpha_1,\alpha_2\rangle=
\langle\alpha_2,\alpha_2\rangle=0$ (i.e., one generator comes from
each ${\bf H}$ factor). The group $Sl(2,\Z)$ may then be taken to act
on this pair to form the required embedding.

Thus far we have built $O(\Lambda^{3,19})\times Sl(2,\Z)$ as part of
the U-duality group. To proceed further we go to another limit point.
In the weak string-coupling limit we expect to recover the conformal
field theory picture of the moduli space. In order to analyze this
limit we use
the root labeled $e_1$ in (\ref{eq:D13}). This leads to the
decomposition
\begin{equation}
\frac{O(5,21)}{O(5)\times O(21)}\cong \frac{O(4,20)}{O(4)\times O(20)}
\times \R_+ \times \R^{24},     \label{eq:N4Bv}
\end{equation}
where the first factor on the right is the moduli space of $N=4$
superconformal field theories corresponding to K3 surfaces
\cite{Sei:K3}, the second factor is the dilaton and the third factor
comes from the 24 R-R fields.
It was shown in \cite{AM:K3p} that the
discrete group action on the first factor should be
$O(\Lambda^{4,20})$.  In the decomposition (\ref{eq:N4Bv}),
the group $O(4,20)$ acts on the space $\R^{24}$ as the group of
rotations with respect to an inner product of signature $(4,20)$.
We will denote this space by $\R^{4,20}$ and note that it forms
the standard vector
representation of the rotation group $O(4,20)$. Another subgroup
of $O(5,21)$ (isomorphic to $\R^{4,20}$) acts as
translations on this space. One can show that the intersection of this
translation group with the discrete group $O(\Lambda^{5,21})$ gives
the group of
 translations  by the lattice $\Lambda^{4,20}$.
By combining these facts with our earlier
decomposition we  find a much larger structure within $O(\Lambda^{5,21})$.

The classical moduli space of K3 surfaces embeds nicely into the
moduli space of $N=4$ conformal field theories showing that
$\Lambda^{3,19}\subset\Lambda^{4,20}$. That is, $\Lambda^{4,20}$ can
be taken to be $\Lambda^{3,19}$ plus one of the two ${\bf H}$
sublattices mentioned earlier on which $Sl(2,\Z)$ acts. Consider
first the generator ``$\tau\to\tau+1$'' of $Sl(2,\Z)$. This generates
translations in $\R^{4,20}$ by one of the unit vectors. Since
$O(\Lambda^{4,20})$ acts within this space, we may generate the whole
$\Lambda^{4,20}$ lattice of translations in $\R^{4,20}$. Now consider
the other generator of $Sl(2,\Z)$, ``$\tau\to-1/\tau$''. This exchanges
the two ${\bf H}$ sublattices of $\Lambda^{5,21}$ that were orthogonal
to $\Lambda^{3,19}$. In fact, it swaps the ${\bf H}$ lattice
``outside'' $\Lambda^{4,20}$ with one of the ones inside. We thus have
three sets of generators to generate a subgroup of
$O(\Lambda^{5,21})$:
\begin{enumerate}
  \item The group $O(\Lambda^{4,20})$.
  \item The group of translations $\Lambda^{4,20}$.
  \item The $\Z_2$ exchanging two ${\bf H}$ sublattices one of which
is the orthogonal complement to $\Lambda^{4,20}$.
\end{enumerate}
This is precisely equivalent to the situation studied in \cite{AM:K3p}
in which the space of conformal field theories on K3 surfaces was analyzed.
In the latter case the three sets of generators understood
``classically'' were
\begin{enumerate}
  \item The group $O(\Lambda^{3,19})$ of classical automorphisms of a K3
surface.
  \item The group of translations $\Lambda^{3,19}$ of the $B$-field.
  \item The $\Z_2$ exchanging two ${\bf H}$ sublattices one of which
is the orthogonal complement to $\Lambda^{3,19}$ (up to an
$O(\Lambda^{3,19})$ rotation) --- namely {\it mirror symmetry}.
\end{enumerate}
Just as in this latter case, the generators we have specified
within $O(\Lambda^{5,21})$ are sufficient to
generate the entire group. That is, the U-duality group for IIB
superstrings on a K3 surface is $O(\Lambda^{5,21})$ as stated in
\cite{W:dyn}.

Note the structure of the R-R part of the moduli space we have
uncovered without really thinking directly about these fields. Firstly
we have shown that they are periodic thanks to the division by the
$\Lambda^{4,20}$ lattice (that is, they live on a 24-dimensional torus).
Secondly we see that these fields transform as a
vector of $O(4,20)$ and so transform in the cohomology
of the K3 surface. We will have more to say about these points later.


\section{The Case {\rm K3}$\times T^2$}

The example we wish to study in detail
concerns a type II superstring compactified to
four dimensions on the product of a K3 surface and a 2-torus. For
definiteness we will first
specify that we have a type IIB string. This gives an $N=4$
supergravity theory in four dimensions with 22 vector multiplets. The
results of
\cite{deRoo:}
tell us that for the moduli space in question $G\cong O(6,22)
\times Sl(2)$. We now wish to study limits in the moduli space which
may allow us to identify parts of the discrete group $U$.

The group $G\cong O(6,22)\times Sl(2)$ has
Dynkin diagram
\begin{equation}
\setlength{\unitlength}{0.007in}%
\begin{picture}(450,70)(95,655)
\put(615,670){\makebox(0,0)[lb]{$e_1$}}
\put(575,670){\makebox(0,0)[lb]{$e_2$}}
\put(535,670){\makebox(0,0)[lb]{$e_3$}}
\put(495,670){\makebox(0,0)[lb]{$e_4$}}
\put(455,670){\makebox(0,0)[lb]{$e_5$}}
\put(415,670){\makebox(0,0)[lb]{$e_6$}}
\put(375,670){\makebox(0,0)[lb]{$e_7$}}
\thinlines
\put(140,690){\circle{10}}
\put(180,690){\circle{10}}
\put(220,690){\circle{10}}
\put(260,690){\circle{10}}
\put(300,690){\circle{10}}
\put(340,690){\circle{10}}
\put(540,690){\circle*{10}}
\put(580,690){\circle*{10}}
\put(620,690){\circle*{10}}
\put(380,690){\circle*{10}}
\put(420,690){\circle*{10}}
\put(100,720){\circle{10}}
\put(500,690){\circle*{10}}
\put(100,660){\circle{10}}
\put(460,690){\circle*{10}}
\put(145,690){\line( 1, 0){ 30}}
\put(185,690){\line( 1, 0){ 30}}
\put(225,690){\line( 1, 0){ 30}}
\put(265,690){\line( 1, 0){ 30}}
\put(305,690){\line( 1, 0){ 30}}
\put(345,690){\line( 1, 0){235}}
\put(104,716){\line( 4,-3){ 32}}
\put(104,663){\line( 4, 3){ 32}}
\end{picture}   \label{eq:D14}
\end{equation}

Let us now identify some of the limits we can obtain. Firstly let us
go to the large area limit of the torus. This is achieved by
decomposing over $e_2$:
\begin{equation}
  \frac{O(6,22)}{O(6)\times O(22)}\times\frac{Sl(2)}{U(1)}\cong
  \frac{O(5,21)}{O(5)\times O(21)}\times\frac{Sl(2)}{U(1)}
  \times\R_+\times\R^{26}.
\end{equation}
The first factor is clearly the space of type IIB strings on a K3
surface as we might expect, the second factor is the complex structure
of the torus and the third factor is its area. The group
$O(\Lambda^{5,21})$ acts on the first factor as explained above and
$Sl(2,\Z)$ acts on the second factor as is well-known.

Now let us go to the weak string-coupling limit. This is achieved by
decomposing over $e_3$:
\begin{equation}
  \frac{O(6,22)}{O(6)\times O(22)}\times\frac{Sl(2)}{U(1)}\cong
  \frac{O(4,20)}{O(4)\times O(20)}\times\frac{Sl(2)}{U(1)}
  \times\frac{Sl(2)}{U(1)}
  \times\R_+\times\R^{49}.      \label{eq:De3}
\end{equation}
The first factor is the moduli space of conformal field theories on a
K3 surface, the second and third factors give the space of conformal
field theories on a torus and the fourth factor is the dilaton. The
groups $O(\Lambda^{4,20})$, $Sl(2,\Z)$ and $Sl(2,\Z)$ are known to
act respectively on the first three factors. We may use the same trick
as in the previous section to
use these two limits to build the U-duality group. The result is
$U\cong O(\Lambda^{6,22})\times Sl(2,\Z)$, in agreement with the
conjecture in \cite{HT:unity}. Again, no larger group than this may act
on the ``Teichm\"uller space'' $G/\com{G}$
 without destroying some nice properties of
the moduli space.
Note that the $Sl(2,\Z)$ which appears as a separate factor in the
U-duality group comes
from the moduli space of complex structures on the torus.

It is interesting to compare this with the moduli space of heterotic
strings compactified toroidally to four dimensions.
In fact, that compactified weakly-coupled heterotic string is the
limiting theory corresponding to the
decomposition over $e_1$.  From our analysis we recover the
precise structure  expected from string-string
duality, except that the $Sl(2,\Z)$ now acts on the four-dimensional
axion-dilaton system. This model was thoroughly analyzed in
\cite{Sen:4d}. An exchange of the r\^oles of $Sl(2,\Z)$
groups between the type II superstring and the heterotic string in
four dimensions was discussed in \cite{Duff:S,W:dyn} although the
two $Sl(2,\Z)$'s exchanged there were not the same ones we have here.
 To understand the
picture fully we need to look at mirror symmetry.

Consider the superconformal field theory description of a $d$-dimensional
Minkowski space $\R^{d-1,1}$, where $d$ is even. For simplicity we go
to the light-cone
gauge in which the target space fermions transform in spinor
representations of
$SO(d-2)$. The superconformal field theory has an
$N=2$ structure where the $\hat{u}(1)$ affine algebra from the $N=2$ algebra
lies in the affine algebra $\widehat{so}(d-2)$.

Let $C$ be a $\Z_2$ automorphism of the weight space of $so(d-2)$ such
that $C:\lambda\to-\lambda$ for any $\lambda$ in the weight space. If
$R$ is a spinor representation of $so(d-2)$ then $C$ is a symmetry of
the corresponding set of weights if and only if $d\in4\Z+2$. If on the
other hand $d\in4\Z$, the action $C$ takes each spinor to the spinor of
opposite chirality (the spinor is in a {\em complex\/}
representation). Note that $C$ changes the sign of the $U(1)$
charge of the weights for any embedding $U(1)\subset SO(d-2)$.

The mirror map takes an $N=(2,2)$ superconformal field theory and
changes the sign of the $U(1)$ charges for the left sector and leaves
the right sector untouched. Thus, for spinors in $d$-dimensional
Minkowski space it will leave everything invariant if $d\in4\Z+2$ and
change the relative chirality of the left and right spinors if
$d\in4\Z$.

Therefore, in $d=10$, the type IIA string is self-mirror and the type
IIB string is self-mirror. Indeed mirror symmetry had better not
identify these two theories since their moduli spaces are quite
different. Similarly the IIA and IIB string are quite different when
compactified on a K3 surface. However, when we compactify on a space
of 2 or 6 real dimensions we expect that the type IIA string
compactified on such a space should be mirror to the IIB string compactified
on the mirror.\footnote{The similarities between IIA and IIB theories
on Calabi--Yau threefolds were  studied some time ago
\cite{CFG:II}, and the connection to mirror symmetry has been pointed
out  in \cite{dWvP:quat,Str:con}.}
 For the 2-torus the mirror map is a special kind of
``$R\leftrightarrow1/R$'' duality and this mirror statement between IIA and IIB
string compactifications was effectively shown in
\cite{DHS:IIAB,DLP:IIAB}.
For compactifications on \CY\ threefolds this statement is potentially
more powerful.  We will discuss mirror symmetry in more detail in section
\ref{s:integral}.

We may apply mirror symmetry to the decomposition (\ref{eq:De3}). The
mirror map acts within the group $O(\Lambda^{4,20})$ for a K3 surface
\cite{AM:K3p} but for the torus it exchanges the two $Sl(2)/U(1)$
factors \cite{DVV:torus} and hence the corresponding $SL(2,\Z)$'s.
Thus our statements regarding the IIB string
compactified on a
K3 times a torus are precisely the same as those for the IIA string so
long as we reverse the r\^oles of the $Sl(2)/U(1)$ factors
and the $Sl(2,\Z)$'s which act upon them.
The moduli space of type IIA strings has a U-duality group
$O(\Lambda^{6,22})\times Sl(2,\Z)$ where now the $Sl(2,\Z)$ acts on
the complex K\"ahler form of the torus. Thus we see a kind of
``triality'' in which this $Sl(2,\Z)$ has three different r\^oles
according to whether we are talking about the IIA, IIB or heterotic string.


\section{Exotic Limits}

It is interesting to explore the other possible decompositions of the
diagram (\ref{eq:D14}). For $e_4$ we obtain
\begin{equation}
  \frac{O(6,22)}{O(6)\times O(22)}\times\frac{Sl(2)}{U(1)}\cong
  \frac{O(3,19)}{O(3)\times O(19)}\times\frac{Sl(3)}{SO(3)}\times
  \frac{Sl(2)}{U(1)}
  \times\R_+\times\R^{69}.
\end{equation}
The first factor is the space of Ricci-flat metrics on a K3 surface of
fixed volume. The second factor is the moduli space of a 3-torus
$T^3$. The third factor we associate to a four-dimensional
axion-dilaton system.
{}From \cite{W:dyn} we thus see that this is eleven-dimensional
supergravity compactified down to four dimensions on K3$\times
T^3$. We have thus been able to identify limits corresponding to (at
least as far as their low-energy effective theories are concerned) type
IIA and type IIB superstrings, the heterotic string and eleven-dimensional
supergravity all in the same moduli space!

Further interesting features are found at other limit points.
Consider $e_5$. This
gives
\begin{equation}
  \frac{O(6,22)}{O(6)\times O(22)}\times\frac{Sl(2)}{U(1)}\cong
  \frac{O(2,18)}{O(2)\times O(18)}\times\frac{Sl(4)}{SO(4)}\times
  \frac{Sl(2)}{U(1)}
  \times\R_+\times\R^{86}.
                \label{eq:De5}
\end{equation}
The first factor appears to be related to the moduli space of K3
surfaces. The space $O(3,19)/(O(3)\times O(19))$ may be thought of as
the Grassmanian of space-like 3-planes in $\R^{3,19}$. To identify
this as the space of metrics on a K3, we take $\R^{3,19}$ to be
$H^2({\rm K3},\R)$ equipped with the natural metric from the cup
product; the 3-plane is spanned by the real and imaginary parts of
the holomorphic 2-form $\Omega$ and the K\"ahler form $J$
\cite{Beau:K3}.
Usually one normalizes $J$ such that $\langle J,J\rangle=1$ which is
equivalent to fixing the volume of the K3 surface to 1 since
\begin{equation}
\Vol({\rm K3}) = \int J\wedge J = \langle J,J\rangle.
\end{equation}
The volume of
the K3 surface may then be represented by a separate $\R_+$ factor in
the moduli space.
Embedded in the space $\R^{3,19}$ we have
the lattice $H^2({\rm K3},\Z)\cong\Lambda^{3,19}$. On the boundary of
this moduli space, the 3-plane will acquire a null direction which
we assume to be along a generator of an ${\bf H}$ sublattice of
$H^2({\rm K3},\Z)$. This gives a natural embedding of
$O(2,18)/(O(2)\times O(18))$ in the boundary of the moduli space of K3
surfaces. It would thus appear natural to identify the first factor
on the left-hand side of (\ref{eq:De5}) with some degenerated K3
surface.

Let us try to identify such a degeneration. We are free to specify
that the null direction in the 3-plane is given by the $J$-direction
without any loss of generality. Note however that this means that we
cannot use the usual normalization conventions of $\langle
J,J\rangle=1$ since now we have $\langle J,J\rangle=0$. Thus the
volume of the limiting K3 surface is zero. The volume element on a smooth
surface is an everywhere positive quantity and so this degeneration of
the metric we are studying must have caused the effective volume to
shrink down to zero at every point in the K3 surface. That is, the
effective dimension of the space must have decreased. Since we are
studying a generic degeneration of this type it seems reasonable to
expect that  this ``squashed'' K3 surface is now an object of real
dimension 3.

The second factor is the moduli space of a 4-torus. We therefore claim
that this limit in the moduli space represents eleven-dimensional
supergravity compactified down to four dimensions by compactifying on a
squashed K3 surface times $T^4$. Recall however that the string theory
in question could also be
considered as the heterotic string compactified on a 6-torus which
is $T^4\times T^2$. By
taking the large radius limit of these $T^4$'s we see that the heterotic
string compactified on a 2-torus is ``equivalent'' (in suitable limits) to
eleven-dimensional
supergravity compactified on the squashed K3 surface. Actually we
could have also seen this more directly by
applying a similar construction to the
observation in \cite{W:dyn} that eleven-dimensional supergravity
compactified on a K3 surface appears to be ``equivalent'' to the
heterotic string on a 3-torus.

The decomposition on $e_6$ is similar. Now we have supergravity
compactified on a K3 surface squashed down to 2 dimensions times a
5-torus. It can be shown that in this case the degeneration in
question may be achieved through a deformation of complex
structure. This allows algebro-geometric techniques to be employed
\cite{FM:BGOD}.
The typical degeneration in algebraic geometry is to a limiting complex
surface with two components which meet on an elliptic curve (i.e., $T^2$),
along which there are 16 singular points.
To get a complete picture of the degeneration it is
still necessary to carefully consider the way in which the Ricci-flat
metric affects the geometry near the degeneration, which we have not
done here. However,
combining the algebro-geometric analysis with our expectation that the
limit is essentially a 2-dimensional object suggests that
the limit can be regarded as roughly being a 2-torus with 16 special
points.  We thus claim that eleven-dimensional supergravity compactified on
this object should be ``equivalent'' to the heterotic string
compactified on a circle.

The decomposition on $e_7$ is perhaps the most interesting one.
The first factor in the decomposition is
$O(16)/O(16)$ which is of course trivial. Note however that this
decomposition may be done in two ways.
When we reduce the K3 surface moduli space in a
degeneration we are effectively removing some ${\bf H}$ sublattices
from the $H^2({\rm K3},\Z)$ lattice.
In the present case we need to remove three ${\bf H}$
sublattices and there are two choices, depending on whether the
 16-dimensional even self-dual lattice
we leave behind is
$\Lambda^8\oplus\Lambda^8$ or the Barnes-Wall lattice $\Lambda^{16}$.
The second factor will be the moduli
space of 6-tori. These two decompositions thus represents ways of
squashing a K3 surface down to one of two 1-dimensional structures
$\Xi_1$ or $\Xi_2$ and the theory in question is now eleven-dimensional
supergravity compactified on a 6-torus times $\Xi_i$.
We claim that the heterotic string in ten dimensions should be
``equivalent'' to
eleven-dimensional supergravity compactified on $\Xi_i$!

The two choices of $\Xi_1$ and $\Xi_2$ for compactifying the
supergravity theory will presumably lead to the $E_8\times E_8$
heterotic string and the Spin(32)$/\Z_2$ heterotic string,
respectively. Clearly the $\Xi_i$ spaces are not manifolds since the
only compact 1-dimensional manifold is a circle and eleven-dimensional
supergravity on a circle is expected to give the type IIA
superstring \cite{T:11d,W:dyn}. The $\Xi_i$ spaces are probably
1-skeletons of some polytopes. It would be interesting, but perhaps
rather difficult, to establish their shape.


\section{Integral Structures}   \label{s:integral}

Now let us consider some general facts about the moduli space of type
II string theories. As discussed above, when we decomposed the space
of type IIB strings on a K3 surface by taking the large dilaton limit,
we were able to identify the R-R fields as living on a 24-dimensional
torus. This torus can also be identified as $H^*({\rm
K3},\R)/H^*_Q({\rm K3},\Z)$. The group $H^*_Q({\rm K3},\Z)$ is the
lattice of integral {\em quantum\/} cohomology which
 coincides with the group $H^*({\rm K3},\Z)$ in some large radius
limit. Away from such a limit, the generators of $H^*_Q({\rm
K3},\Z)$ cannot be associated with pure $p$-forms but will have other
degrees mixed in. The precise way in which this happens is known from
conformal field theory and mirror symmetry \cite{AM:K3p}.

A similar phenomenon can also be deduced from the above calculation for
a IIA superstring compactified on a K3 times a torus. In the weak
coupling limit we have
\begin{equation}
  \frac{O(6,22)}{O(6)\times O(22)}\times\frac{Sl(2)}{U(1)}\cong
  \frac{O(4,20)}{O(4)\times O(20)}\times\left(\frac{Sl(2)}{U(1)}\right)_1
  \times\left(\frac{Sl(2)}{U(1)}\right)_2
  \times\R_+\times\R^{49},      \label{eq:De3x}
\end{equation}
where the $(Sl(2)/U(1))_2$ factor is to be identified with the
$Sl(2)/U(1)$ on the left and so provides the ``area+$B$-field'' degrees of
freedom of the torus on which we are compactifying. Analysis of
the decomposition (\ref{eq:De3x}) shows that the group $O(4,20)\times
Sl(2)_1$ acts
on the $\R^{49}$ space through a representation of the form
$({\bf 24},{\bf 2})+({\bf 1},{\bf 1})$. Let us concentrate on the
$\R^{48}$ subspace of R-R fields forming the nontrivial irreducible
representation (the other field is simply the four-dimensional
axion that was partnered with the dilaton). From our knowledge of the K3
{\em quantum\/} cohomology and
the fact that
the group $Sl(2,\Z)_1$ exchanges 1-cycles on the torus we see
that $\R^{48}\cong H^{\rm odd}({\rm K3}\times T^2,\R)$. In the large
radius limit of the compact space this is consistent with the counting
of the R-R fields coming from the reduction of the 1-form and 3-form
R-R fields of the ten-dimensional type IIA string. The knowledge of the
complete U-duality group for this case also allows us to find the
discrete identifications on the R-R moduli space. The result is that
the R-R moduli space becomes
\begin{equation}
  \cM_A\cong \frac{H^{\rm odd}(X,\R)}{H^{\rm odd}_Q(X,\Z)},
                \label{eq:RRA}
\end{equation}
where $X$ is the space on which are compactifying.
Similarly for the IIB string we obtain the result that the 48 R-R
scalars parameterize
\begin{equation}
  \cM_B\cong \frac{H^{\rm even}(X,\R)}{H^{\rm even}_Q(X,\Z)}.
                \label{eq:RRB}
\end{equation}

We conjecture that this structure survives in the case that $X$ is a
more general \CY\ manifold. We should  add some cautionary notes
to this statement however. The situation is best compared to that of
the r\^ole the $B$-fields play in the moduli space of $N=2$
superconformal field theories. In terms of the non-linear \sm\ with
target space $X$, it is
easy to see that the $B$-fields parameterize the space
$H^2(X,\R)/H^2(X,\Z)$. In terms of the
exact conformal field theory one sees this by putting a system of
``flat'' coordinates in the neighbourhood of the large radius
limit. This picture is only natural in the \CY\
phase of the theory (in the sense of \cite{W:phase,AGM:II}). In other
phases one gets quite different pictures.
The same is probably true of the R-R system we are
considering here and we only have a right to expect the behaviour we
claim in some neighbourhood of weak string-coupling. In particular our
conjecture is not incompatible with the local analysis of the moduli
space done in \cite{CFG:II,FS:quat} where it was discovered that the R-R
fields
 along with the axion-dilaton system parameterize a space locally
of the form $SU(1,n)/(U(1)\times SU(n))$.
Let us also note that the full U-duality picture is probably less useful in
this general \CY\ case in the same way that T-duality becomes less
useful in this more general case (see, for example \cite{me:N2lect}).

In our picture the moduli space of
compactifications of type II superstrings has the form of a fibration
where the base space is the moduli space of $N=2$ superconformal field
theories given by NS-NS moduli (including the axion-dilaton) and the
fiber is a torus of the form
(\ref{eq:RRA}) or (\ref{eq:RRB}).
Each fiber also has a natural complex structure making it into the
``intermediate Jacobian'' of the Calabi--Yau manifold, and the total
space of the fibration is well-understood \cite{DM:cubics}.

If we consider $X$ at large radius
limit, then simple dimensional reduction of the 1-form and 3-form of
the IIA superstring and 0-form, 2-form and ``anti-self-dual'' 4-form
of the IIB superstring tell us that at least the dimension counting in
equations (\ref{eq:RRA}) and (\ref{eq:RRB}) is correct.
Assuming that $h^{1,0}(X)=0$ which is a reasonable assumption for a
generic \CY\ manifold, $H^3(X)$ provides all the odd cohomology and so
$H^{\rm odd}_Q(X,\Z)\cong H^3(X,\Z)$. Thus we may move away from the large
radius limit of $X$ and still understand the denominator of
(\ref{eq:RRA}).

Given a mirror pair $X$ and $Y$, part of the mirror map conjecture (which has
yet to be proven in full
generality) is that the ``horizontal'' integral structure
$H^3(Y,\Z)$ should somehow be equivalent to the ``vertical'' integral
structure $H^0(X,\Z)\oplus H^2(X,\Z)\oplus H^4(X,\Z)\oplus H^6(X,\Z)$
in the large radius limit \cite{AL:qag,Mor:Gid}.\footnote{A very precise
version of this conjecture including a
specific prediction concerning
the mirror map has been formulated in \cite{Mor:pred},
and the conjecture was checked in several examples.}
If this is
true, we may determine the moduli space of R-R fields in the IIB
superstring compactification by using the fact that $H^{\rm
even}_Q(X,\Z) \cong H^{\rm odd}_Q(Y,\Z)$.

The assumption of S-duality for the IIB superstring goes
some way towards proving our conjecture about the moduli space. The
ten-dimensional axion, $a$, in the R-R sector is associated to
$H^0(X)$ when $X$ is at large radius. The assumption of S-duality for
the type IIB string gives
$a\cong a+1$. This part of the moduli space may be rather trivially
identified as $H^0(X,\R)/H^0(X,\Z)$. When $X$ is at large radius $Y$ will be at
``large complex structure'', i.e., it will be degenerating. The
direction $H^0(X)$ is ``mirror'' to the direction $H^{3,0}(Y)$ in this
limit \cite{Cand:mir}. In this way, the axion periodicity for the IIB
superstring corresponds to an element of $H^3(Y,\Z)$ for the IIA
compactification.
Now we may consider the
monodromy of $H^3(Y,\Z)$ as we move around the moduli space of complex
structures of $Y$. This will take the single element we have and
produce more elements. In fact we can generate  enough elements this way
to build at least a finite index sublattice of $H^3(Y,\Z)$.
(The fact that we do not necessarily generate the whole
$H^3(Y,\Z)$ means that our argument falls short of a full proof of the
conjecture.)
Taking the mirror map to go back to $X$ we see
that we are naturally forced to consider not only shifts in $H^0(X)$
but all of the even cohomology of $X$.

It is interesting to note that this picture of the moduli space of
type II compactifications cannot be seen purely in terms of
world-sheet (i.e., conformal field theory) considerations or target
space effective field theory considerations (i.e., explicit
compactification down from ten to four dimensions). The R-R sector of the
moduli space is rather trivial in the conformal field theory point of
view since all the fields there do not couple to any R-R charges. On
the other hand the target space point of view knows only about
classical cohomology and cannot describe a shift in quantum
cohomology unless one is near the large radius limit where such
objects can be identified with classical forms of definite degree. It
would appear that we are in some way probing the effects of
nonperturbative phenomena in the, as yet unknown, correct formulation
of string theory.


\section*{Acknowledgements}

We thank B.~Greene and R.~Plesser for useful conversations.
The work of P.S.A. is partially supported by a grant from the National
Science Foundation,
and that of D.R.M. by the United States Army
Research Office through the Mathematical Sciences Institute of
Cornell University, contract DAAL03-91-C-0027,
and  by the National Science Foundation through
grants DMS-9401447 and PHY-9258582.


\end{document}